\documentclass{appolb}
\usepackage[utf8]{inputenc}
\usepackage[T1]{fontenc}
\usepackage{color,graphicx}
\usepackage{wrapfig}
\usepackage{booktabs}
\usepackage{epstopdf}
\begin{document}
\eqsec  
\title{Control of transport characteristics in two coupled Josephson junctions}%

\author{J. Spiechowicz, L. Machura, M. Kostur, J. {\L}uczka 
\address{Institute of Physics, University of Silesia, 40-007 Katowice, Poland}
}

\maketitle
\begin{abstract}
We report on a theoretical study of transport properties of two coupled  Josephson junctions and 
compare two scenarios for controlling the current-voltage characteristics when the system is driven 
by an external biased DC current and unbiased AC current consisting of one harmonic. In the first 
scenario, only one junction is subjected to both DC and AC currents. In the second scenario the signal
is split -- one junction is subjected to the DC current while the other is subjected to the AC current. 
We study DC voltages across both junctions and find diversity of anomalous transport regimes for 
the first and second driving scenarios.  

\end{abstract}
\PACS{05.60.-k, 74.50.+r, 85.25.Cp 05.40.-a}
  
\section{Introduction}
\label{sec1}
In symmetric devices, transport can be generated by nonequilibrium forces which can break space 
or time symmetry. In mechanical systems like movement of a Brownian particle in a spatially 
periodic and symmetric potential, the directed motion can be induced by an external static load 
forces or by unbiased multi-harmonic forces. Another class of systems with broken spatial symmetries 
is related to ratchet systems studied intensively during last twenty years \cite{ratchets}. 
In the paper, we study 
a relatively simple symmetric system which is constructed from the well-known physical elements: 
Josephson junctions. Their role in physics is invaluable and multifaceted, 
offering a rich spectrum of beneficial applications: from the definition
of the voltage standard, through more practical devices as elements in high
speed circuits \cite{barone}, to the future applications in quantum computing devices \cite{qcjj}. 
We study two Josephson junctions coupled by an external resistance. The evolution of the system 
can manifest counterintuitive 
nature when we test its response to a constant external current: it can exhibit the negative 
resistance \cite{nr}. We can formulate the general question: how to manipulate the system by 
external drivings to get optimal and desired transport behaviour? To answer this question, we propose to manipulate the system by two  combinations of external currents. 

The paper is organized in the following way: In Sec. \ref{sec2}, we define the model 
and  provide all necessary definitions and notation. Next, in Sec. \ref{sec3}, we study the response of the system 
in the case when external DC and AC currents are applied to one junction only. In Sec. \ref{sec4}, we analyze the 
case when the DC current is applied to the first junction and the AC current is applied to the second 
junction. In Sec. \ref{sec5}, comparison of transport characteristics for two driving scenarios is presented.  

\section{Model}
\label{sec2}
\begin{figure}[h]
	\centering
	\includegraphics[width=0.5\linewidth]{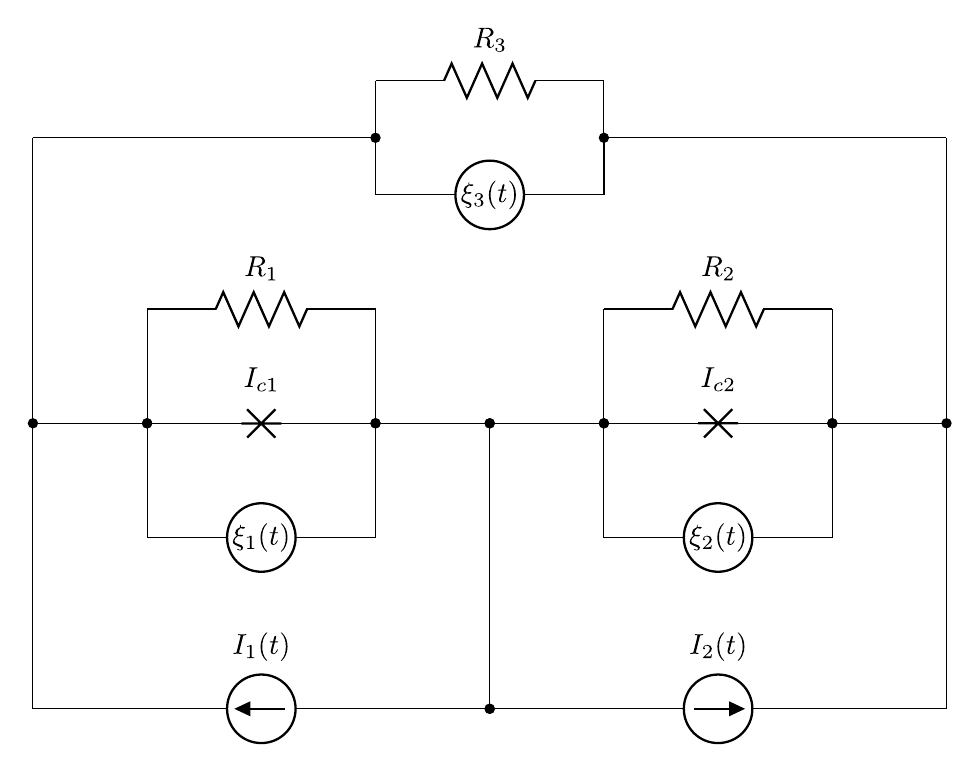}
	\caption{The system of two coupled Josephson junctions shunted by an external resistance $R_{3}$ and driven by the  external currents $I_{1}(t)$ and $I_{2}(t)$.}
	\label{fig1}
\end{figure}

From a more fundamental point of view, we consider a system which consists of two coupled (interacting) 
subsystems  and we want to uncover its transport properties induced by coupling between two subsystems. 
As a particular  example of the real physical structure, we study a Josephson junction device which consists  
of a coupled pair of resistively shunted Josephson junctions 
characterized by the critical Josephson supercurrents $(I_{c1}, I_{c2})$, normal state resistances $(R_{1}, R_{2})$ 
and phases  $(\phi_{1}, \phi_{2})$ \cite{Ner84}. A schematic circuit representing the model  is shown in Fig.  \ref{fig1}. 
The system is externally shunted by the resistance 
$R_{3}$ and driven by two current sources  $I_1(t)$ and $I_2(t)$ acting on the first and second junctions, 
respectively.  
We also include into the model Johnson-Nyquist thermal noise sources 
$\xi_{1}(t), \xi_{2}(t)$ and $\xi_{3}(t)$ associated with the corresponding resistances 
$R_1, R_2$ and $R_3$ according to  the fluctuation--dissipation theorem. 
We assume the semiclassical and small junction regimes where the  spatial dependence of characteristics 
can be neglected and  photon-assisted tunnelling phenomena do not 
contribute to the general dynamics. It is the regime in which the so-called Stewart-McCumber 
model \cite{stewart} holds true. The range of validity of this model is discussed in detail in the review paper \cite{kautz}. 

The Kirchhoff current and voltage laws, and two Josephson relations
allow for the full description of the phase dynamics of both  junctions within assumed restrictions. 
The dimensional form of the equations of motion  is presented in Ref. \cite{JPC_2012}. Therein, the dimensionless variables and parameters are defined, and the dimensionless form of dynamics is presented. 
Here, we recall only  the dimensionless version of equations of motion for the phases $\phi_1=\phi_1(\tau)$ and $\phi_2=\phi_2(\tau)$, namely,  
\begin{eqnarray}
  \label{fi}
    \dot{\phi}_1 = I_1(\tau)- I_{c1} \sin \phi_1 
      +\alpha [I_2(\tau)  -  I_{c2} \sin \phi_2] + \sqrt{D}\; \eta_1(\tau),  \nonumber\\
    \dot{\phi}_2 =  \alpha \beta[I_2(\tau) -  I_{c2} \sin \phi_2]
    	       + \alpha [ I_1(\tau) - I_{c1} \sin \phi_1] + \sqrt{\alpha \beta D} \;\eta_2(\tau),    
\end{eqnarray}
where the dot denotes a derivative with respect to dimensionless time $\tau$, which is defined by the dimensional time $t$ in the 
following way \cite{Ner21}
\begin{eqnarray}
  \label{time}
	\tau = \frac{2eV_0}{\hbar} t, 
\end{eqnarray}
where 
\begin{eqnarray}
  \label{voltage}
	V_0 =  I_c \frac{R_1(R_2+R_3)}{R_1+R_2+R_3}, \quad I_c=\frac{I_{c1}+I_{c2}}{2} 
\end{eqnarray}
are the characteristic voltage and averaged critical supercurrent, respectively. 
All dimensionless currents in Eqs. (\ref{fi}) are in units of $I_c$. E.g., $I_{c1} \to I_{c1}/I_c$.  The parameters 
\begin{eqnarray}
  \label{alfa}
  \alpha = \left(1 + \frac{R_3}{R_2}\right)^{-1}  \in [0, 1], \quad \beta = 1 + \frac{R_{3}}{R_{1}}. 
\end{eqnarray}
 We assume that all resistors are at the same temperature $T$ and that the noise 
sources are modelled by independent $\delta$--correlated zero-mean Gaussian white noises 
$\xi_{i}(t)\ (i = 1,2,3)$, i.e.,  
$\left< \xi_{i}(t)\xi_{j}(s)\right> = \delta_{ij}\delta(t - s)$ for $i, j \in \{1, 2, 3\}$. 
The $\delta$--correlated zero-mean Gaussian white noises   $\eta_1(t)$ and $\eta_2(t)$ which appear 
in  Eqs. (\ref{fi}) are linear combinations of the original noises $\xi_{i}(t)$ and the resulting 
dimensionless noise strength reads $D = 4ek_{B}T/\hbar I_{c}$.

Here we would like to stress out that the dimensional equations of motion for phases 
$\phi_1$ and $\phi_2$ are symmetrical with respect to the transformation $R_1\leftrightarrow R_2$. 
However, their dimensionless equivalents (\ref{fi}) are not symmetrical with respect to the change 
$R_1\leftrightarrow R_2$. It is because of: (i) the definition of the dimensionless time (\ref{time}) 
which is extracted from the equation of motion for $\phi_1$ and, in consequence, (ii) the asymmetry 
of $V_0$ in Eq. (\ref{voltage}) with respect to $R_1$ and $R_2$ . 

Sometimes, it might be helpful to image the dynamics of two Josephson junctions described by Eqs. 
(\ref{fi}) as a motion of two interacting Brownian particles driven by external 
time-dependent forces. In this mechanical framework we have the following correspondence:  
$x_{1} = \phi_{1}$, $x_{2} = \phi_{2}$, where $x_{i}$ for $i = 1,2$ stands for the  
coordinate of the first and second particles, respectively.
The main transport characteristic of such a mechanical system are the long-time averaged velocities 
$v_1 = \langle\dot{\phi_{1}}\rangle$ and $v_2 = \langle\dot{\phi_{2}}\rangle$ of the first and 
 second particles, respectively. In terms of the Josephson junction system it corresponds to the 
dimensionless long-time averaged voltages $v_1 = \langle\dot{\phi_{1}}\rangle$ and 
$v_2 = \langle\dot{\phi_{2}}\rangle$ across the first and second junctions, respectively 
(from the Josephson relation,  the dimensional voltage $V=(\hbar/2e)d\phi/dt$ and therefore  $d\phi/d\tau = V/V_0$). The 
junction resistance (or equivalently conductance) translates then into the particle mobility. 
The phase space of the deterministic system (\ref{fi}) is three-dimensional, 
namely $\{x_{1} = \phi_{1}, x_{2} = \phi_{2}, x_{3} = \omega t\}$. Note that it is a minimal dimension 
for the system to display chaotic evolution in continuous 
dynamical systems which may be a key feature for anomalous transport to arise 
\cite{KosMac2006, MacKos2007,reim,ANM}. 
Other aspects of dynamics  of two coupled Brownian particles has been studied in literature \cite{hennig}. However, experimental realizations of such systems would be difficult to construct.   

The considered system is characterized by four dimensionless material constants: $\{I_{c1}, I_{c2}, \alpha, \beta\}$ 
and by the dimensionless temperature $D$. Additionally, drivings $I_1(\tau)$ and $I_2(\tau)$ are also 
characterized by some parameters. In order to reduce a number of parameters of the model we consider a 
system of two identical junctions, i.e., $R_{1} = R_{2}$ and $I_{c1} = I_{c2} \equiv 1$. In such a case  
$\alpha \beta = 1$ and Eqs. (\ref{fi}) takes the  symmetric form 
\begin{eqnarray}
  \label{fi2}
    \dot{\phi}_1 = I_1(\tau)-  \sin \phi_1 
      +\alpha [I_2(\tau)  -  \sin \phi_2] + \sqrt{D}\; \eta_1(\tau),  \nonumber\\
    \dot{\phi}_2 =  I_2(\tau) -   \sin \phi_2
    	       + \alpha [ I_1(\tau) -  \sin \phi_1] + \sqrt{ D} \;\eta_2(\tau).     
\end{eqnarray}
The parameter $\alpha$ plays the role of the coupling constant between two 
junctions and can be changed by variation of the external resistance $ R_3$. The set of two differential 
equations is decoupled for $\alpha = 0$ which results with two independent subsystems. It is the case when 
$R_3 \to \infty$. Note that when $R_3 =0$, the parameter $\alpha = 1$ and the system is coupled.


\section{The first scenario: DC and AC currents applied only to one junction}
\label{sec3}

The external dimensionless currents $I_1(\tau)$ and $I_2(\tau)$ can be modelled in a various way. 
In experiments with Josephson junctions, 'the most popular' three classes of currents have been 
applied: DC currents, AC currents consisting one harmonic and AC biharmonic currents:  
\begin{eqnarray}
  \label{I(t)}
	I_i(\tau) = I_i + a_i \cos (\omega_i \tau) + b_i \cos ( \Omega_i \tau + \theta_i), 
	\quad i=1, 2. 
\end{eqnarray}
We start with the first scenario in  which we apply the external current to the first junction only, namely,  
\begin{eqnarray}
  \label{I1(t)}
	I_1(\tau) = I_1 + a_1 \cos (\omega \tau), \quad  I_2(\tau) = 0.
\end{eqnarray}
In this special case  Eqs. (\ref{fi2}) take the form:
\begin{eqletters}
	\label{dimlessid}
	\begin{eqnarray}
	\dot{\phi}_{1} &=& I_{1} - \sin\phi_{1}  - \alpha \sin\phi_{2} + a_{1}\cos(\omega \tau) + \sqrt{D} \;\eta_1(\tau),
	\label{dimlessid1}
	\\
	\dot{\phi}_{2} &=& \alpha I_{1} - \sin\phi_{2}  - \alpha \sin\phi_{1} + \alpha a_{1}\cos(\omega \tau) + \sqrt{D}\;\eta_2(\tau).
	\label{dimlessid2}
	\end{eqnarray}
\end{eqletters}
This case was considered in Ref. \cite{JanLuc2011} in the context of indirect control of 
transport and absolute negative conductance induced by coupling between two junctions.  
Here, for the reader's convenience, we recall the main transport characteristics of the system but 
just before we'll do it let us clarify some technical issues.
 
The above set of equations cannot be handled by known analytical methods  for solving ordinary 
differential equations. For this reason we have carried out extensive numerical simulations. We 
have used the stochastic version of Runge-Kutta algorithm of the $2^{nd}$ order with the time step 
of $10^{-3} \cdot (2\pi/\omega)$. The initial phases $\phi_{1}(0)$ and $\phi_{2}(0)$ have been 
randomly chosen from the interval $[0, 2\pi]$. Averaging was performed over $10^{3} - 10^{6}$ 
different realizations and over one period of the external driving $2\pi/\omega$. Numerical 
simulations have been carried out using CUDA environment on desktop computing processor NVIDIA 
GeForce GTX 285. This gave us possibility to speed up the numerical calculations up to few 
hundreds times more than on typical modern CPUs. Details on this very efficient 
method can be found in \cite{cuda}.

The voltage $v_{i} = v_{i}(I_{1}), i=1,2$,  is typically a nonlinear and non-monotonic function of the 
DC current $I_1$. In the normal transport regime the nonlinear resistance  or the static resistance $R_i = R_i(I_{1}) = v_{i}(I_{1})/I_{1}$ (or equivalently 
conductance $C_i=1/R_i$) is positive at a 
fixed bias $I_1$. When the system 
response is opposite to the external driving, i.e., when  $R_i < 0$ we reveal the anomalous transport regime 
with  absolute negative resistance (ANR)
\cite{MacKos2007,reim} or nonlinear negative resistance (NNR) \cite{ANM}. 

Now we would like to address some general comments about the long-time behaviour of the considered 
system (\ref{dimlessid}). Let us consider the voltages $v_{1} = v_{1}(I_{1})$ and 
$v_{2} = v_{2}(I_{1})$ as  functions of the DC bias.   If we make the transformation $I_{1} \to -I_{1}$ to Eqs.  (\ref{dimlessid}), we note that 
$v_{1}(-I_{1}) = -v_{1}(I_{1})$ and $v_{2}(-I_{1}) = -v_{2}(I_{1})$ (because the functions $\sin\phi_{i}$ and $\cos(\omega \tau)$ are symmetric and noises $\eta_i(\tau)$ are also symmetric). From these relations it 
follows that $v_{1}(0) = -v_{1}(0)$, $v_{2}(0) = -v_{2}(0)$ and we deduce that $v_{1}(0) = 0$ 
and $v_{2}(0) = 0$ when the DC bias is zero, i.e., for  $I_{1} = 0$. 
From results of Ref. \cite{JanLuc2011} it follows that 
for high frequency $(\omega > 5)$, the long time averaged voltages $v_{i}$ are  
 negligible small. It is because very fast positive and negative changes of the driving cannot induce transport. If the 
DC current  is sufficiently large,  it is rather obvious that voltages across both junctions has 
the same sign as the DC bias and depend (almost) linearly on the DC current. 
For the  DC current $I_1 > 0$, one can identify three  remarkable and distinct  transport  regimes:
\begin{itemize}
\item[(I)] $v_1 > 0$ and $v_2 >0$,
\item[(II)] $v_1 > 0$ and $ v_2 <0$,
\item[(III)] $v_1 < 0$ and $ v_2 < 0$.
\end{itemize}
The regime (IV):  $v_1 < 0$ and $ v_2 > 0$ has not been detected. 
\begin{figure}[h]
	\centering
	\includegraphics[width=0.49\linewidth]{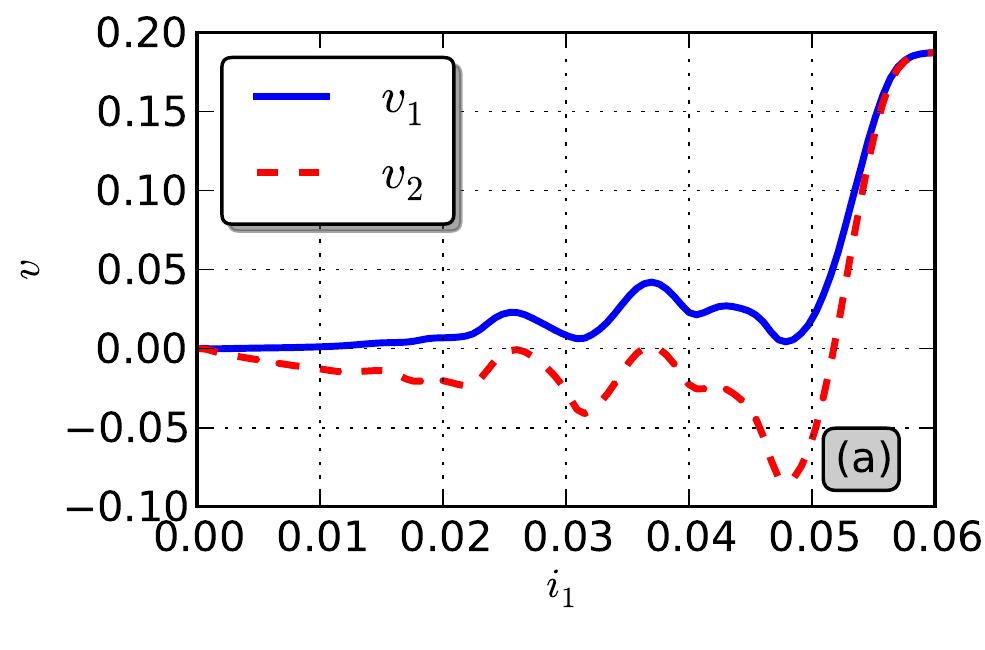}
	\\
	\includegraphics[width=0.49\linewidth]{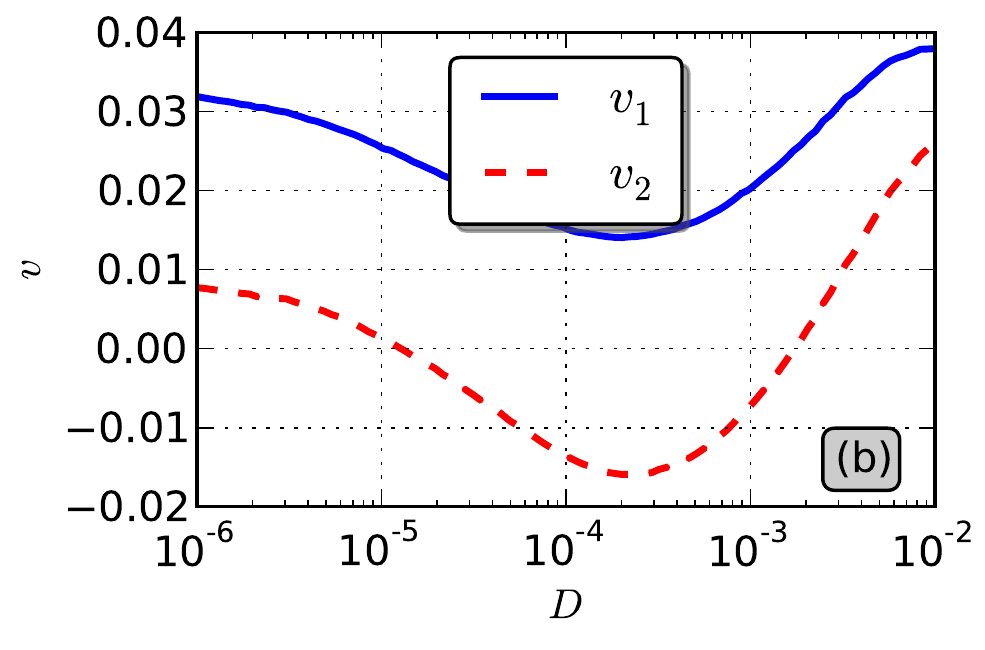}
	\includegraphics[width=0.49\linewidth]{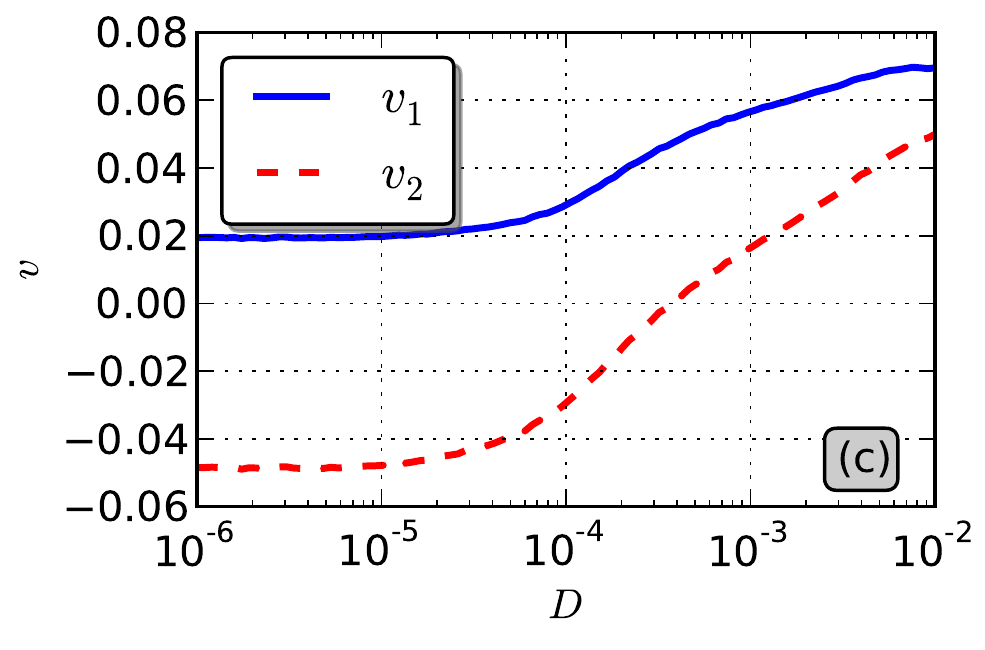}
	\caption{(color online). The first scenario: the long-time averaged voltages $v_{i}$ for $i = 1,2$ of the first and 
	second junction. Panel (a) illustrates the dependence on the DC current $I_{1}$ at the fixed 
	temperature $D = 2 \cdot 10^{-5}$. In the panel (b) and (c) dependence on temperature $D$ is 
	depicted for $I_{1} = 0.025$ and $I_{1} = 0.0455$, respectively. Other parameters read:  
	coupling strength $\alpha = 0.77$, amplitude $a_{1} = 1.775$ and frequency $\omega = 0.1875$ 
	of the AC driving, $I_{2} = a_{2} = 0$.}
	\label{fig2}
\end{figure}
From the analysis reported in Ref. \cite{JanLuc2011} it follows that the most interesting transport effects can take 
place in the regime of small $I_{1}$. Indeed, it has been  found that for $I_{1} < 0.1$ the absolute 
value of the long-time average voltage across both junctions takes its highest values for 
$\omega < 1$. For stronger coupling, strips of non-zero average voltage begin to appear at 
progressively lower values of the amplitude $a_{1}$ of the AC driving. They are also visible for 
the average voltage of the first junction, which means that the strips represent the regimes in the 
parameter space where both junctions operate synchronously. It is illustrated in Fig. \ref{fig2}.  
In this regime of parameters, we can detect several interesting effects: 
 \begin{itemize}
\item the DC voltage $v_1$ is positive but the voltage $v_2$ exhibits ANR for $I_{1} \to 0$ and NNR for larger value of  $I_{1}$. 
\item There are   two 
different mechanisms generating negative resistance in the second junction: 
 \begin{itemize}
 \item  In the case $I_{1} = 0.025$, 
the negative resistance is induced by thermal fluctuations:  for $D=0$ the voltage $v_2 > 0$ and when temperature increases  $v_2$ becomes negative.  There is a restricted interval of 
temperature where the voltage $v_{2}$ is negative.
\item In the case $I_{1} = 0.0455$, the negative 
resistance is generated by the deterministic dynamics because in the deterministic limit (when $D=0$) the DC voltage $v_2 < 0$.  
\end{itemize}
 \end{itemize}
\begin{figure}[h]
	\centering
	\includegraphics[width=0.49\linewidth]{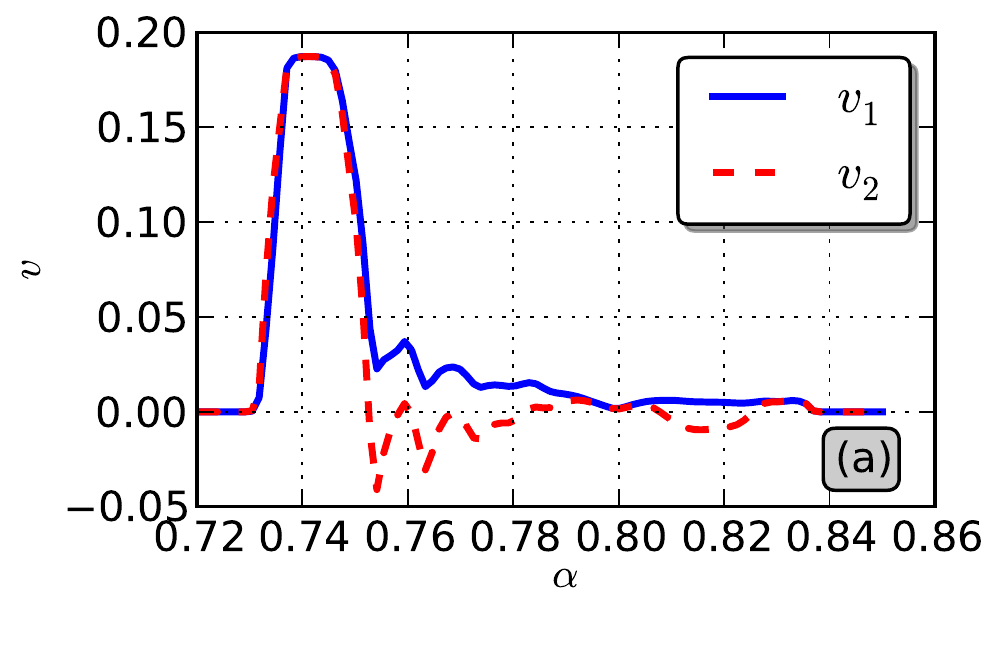}
	\includegraphics[width=0.49\linewidth]{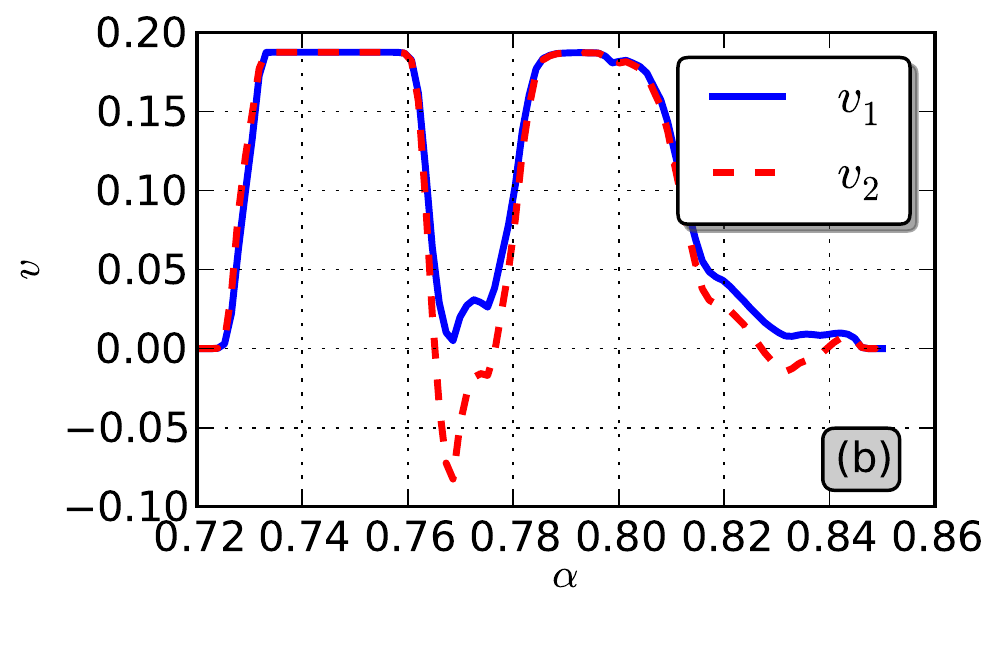}
	\caption{(color online). The first scenario: the long-time averaged voltages $v_{i}$ for $i = 1,2$ across the 
	first and second junction. The influence of the coupling parameter $\alpha$ on transport 
	properties is depicted in the panels (a) and (b) for two fixed values of the DC current 
	$I_{1} = 0.025$ and $I_{1} = 0.0455$, respectively. Other parameters are:  temperature 
	$D = 2 \cdot 10^{-5}$, amplitude $a_{1} = 1.775$ and frequency $\omega = 0.1875$ of the AC 
	driving, $I_{2} = a_{2} = 0$.} 
	\label{fig3}
\end{figure}
The anomalous transport effects like absolute negative resistance  cannot occur in the decoupled system because in this case two  decoupled and independent equations correspond to the overdamped dynamics for which the long-time average $v_1 = \langle \dot{\phi}_{1}\rangle$ has the same sign as $I_1$ 
and $v_2 = \langle \dot{\phi}_{2}\rangle =0$. 
Fig. \ref{fig3} shows how the average voltage across the junctions depends on the coupling constant 
$\alpha$. One can note windows of $\alpha$ for which  anomalous transport  can be 
observed.


\section{The second scenario: DC applied to the first junction and AC applied to the second junction}
\label{sec4}

Nowadays technology allows  experimentalists to apply driving to each of the junctions separately.
In the following we would like to consider the scenario in which the DC current $I_1(\tau) =I_1$  
is applied to  the first junction and the AC current $I_2(\tau)= a_{2}\cos(\omega \tau)$ is applied 
to the second junction. The corresponding dynamics is described by the  special 
case of Eqs.  (\ref{fi2}), namely, 
\begin{eqletters}
	\label{dimlessdiv}
	\begin{eqnarray}
	\dot{\phi}_{1} &=& I_{1} - \sin\phi_{1}  - \alpha \sin\phi_{2} + \alpha a_{2}\cos(\omega \tau) + \sqrt{D}\;\eta_1(\tau),
	\label{dimlessdiv1}
	\\
	\dot{\phi}_{2} &=& \alpha I_{1} - \sin\phi_{2} - \alpha \sin\phi_{1} + a_{2}\cos(\omega \tau)  + \sqrt{D}\;\eta_2(\tau). 
	\label{dimlessdiv2}
	\end{eqnarray}
\end{eqletters}
\begin{figure}[htbp]
	\centering
	\includegraphics[width=0.49\linewidth]{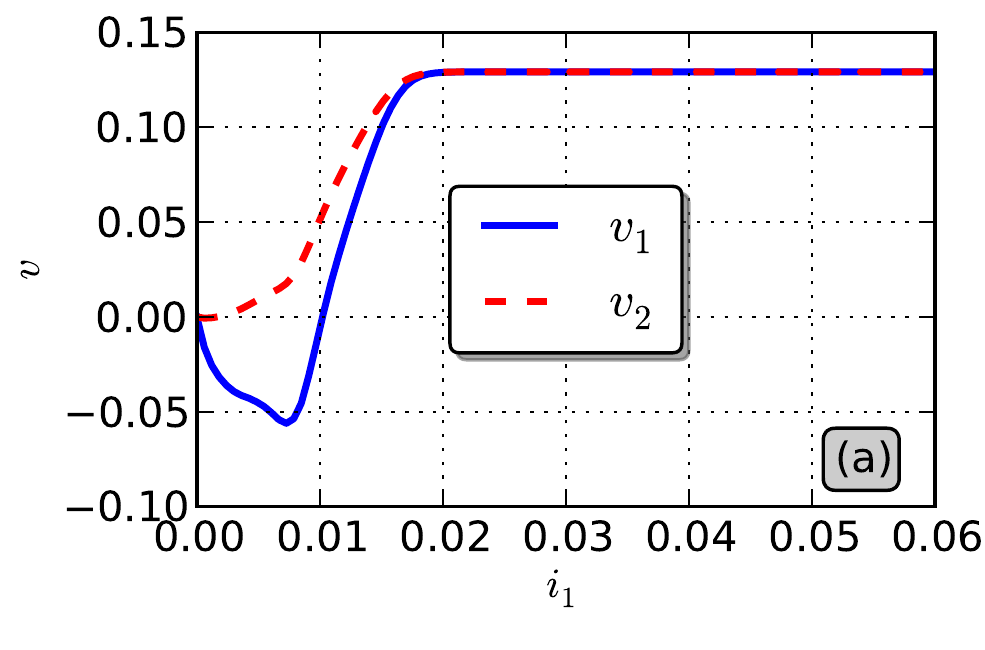}
	\\
	\includegraphics[width=0.49\linewidth]{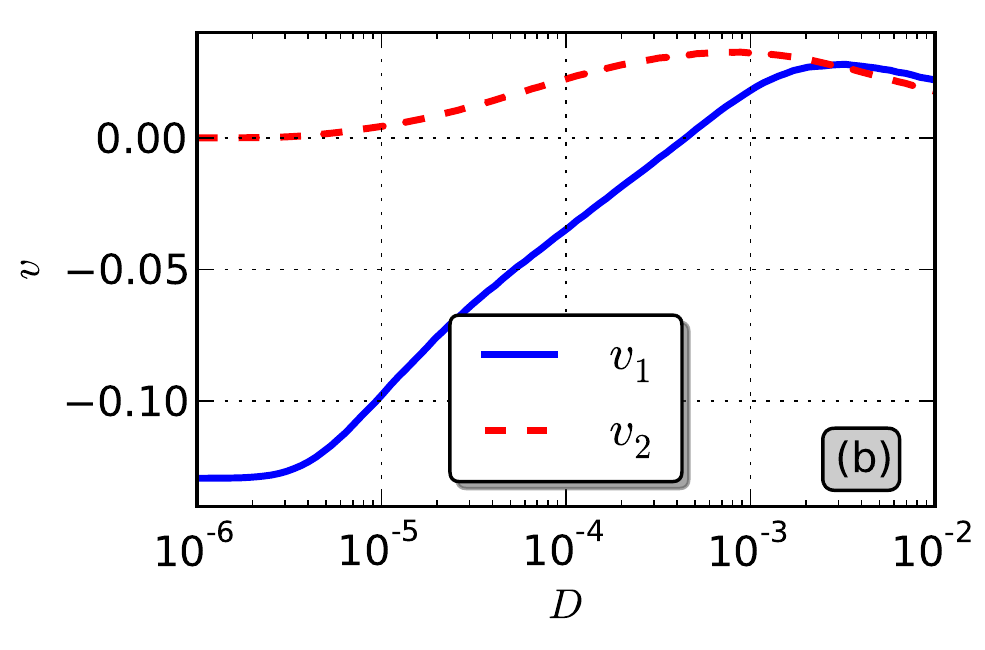}
	\includegraphics[width=0.49\linewidth]{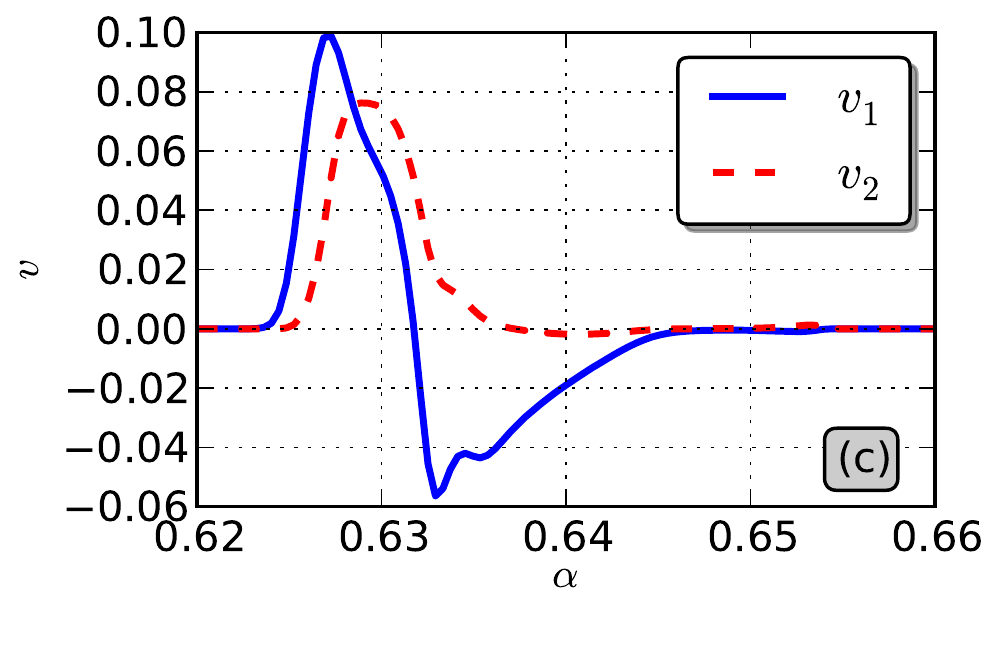}
	\caption{(color online). The second scenario: the long-time averaged voltages $v_{i}$ for $i = 1,2$ across the first and 
	second junction. Panel (a) illustrates the dependence on the DC bias $I_{1}$ at fixed value of 
	amplitude 	$a_2 = 3.0023$, frequency of AC driving $\omega = 0.1292$, the coupling strength 
	$\alpha = 0.6333$ 	and temperature $D = 4.4 \cdot 10^{-5}$. 
	Panel (b) illustrates the temperature dependence for the   DC current $I_{1} = 0.0067$. Other 
	parameters are the same as in panel (a). 	
	Panel (c) shows the role of coupling $\alpha$. Other parameters are the same as in panel (b).}
	\label{fig4}
\end{figure}
This scenario leads to transport characteristics which in general are different than in the first scenario. 
In particular, for the  DC current $I_1 > 0$, one can identify only two transport regimes  where:
\begin{itemize}
\item[(I)] $v_1 > 0$ and $v_2 >0$,
\item[(II)] $v_1 < 0$ and $ v_2 > 0$.
\end{itemize}
The regimes $\{v_1 > 0$ and $ v_2 < 0 \}$ and $\{v_1 < 0$ and $ v_2 < 0\}$ have not been detected. 
This means that for this type of driving the transport properties are a little bit modest. 

The regime (II) seems to be more interesting. In Fig. \ref{fig4}, we present the current-voltage characteristics 
in this regime. The unique feature is the emergence of the absolute negative resistance and the interval of 
$I_{1}$ where the averaged voltage 
across the first junction $v_{1}$ is negative. This is to be contrasted with the 
voltage $v_{2}$ which assumes only positive values. The regime of the nonlinear negative resistance is not 
found in this scenario. The most profound ANR effect occurs for the dc current $I_1 = 0.0067$. 
For this value, in panel (b) of Fig.  \ref{fig4},  we show the voltage dependence on temperature $D$ of 
the system.  A closer inspection of the panel (b) of Fig. \ref{fig4} 
reveals a mechanism responsible for generating of anomalous transport. The negative resistance is solely induced by  
deterministic dynamics and even at  zero temperature $D = 0$ the resistance is negative. For this 
chaotic--assisted mechanism, temperature plays destructive role: if  temperature increases the 
effect disappears and for temperature  $D$ greater  than 
$D = 5 \cdot 10^{-4}$ the averaged voltage $v_{1}$ is positive.
 
\section{Comparison of transport characteristics for two scenarios} 
\label{sec5}
\begin{figure}[htbp]
	\centering
	\includegraphics[width=0.49\textwidth]{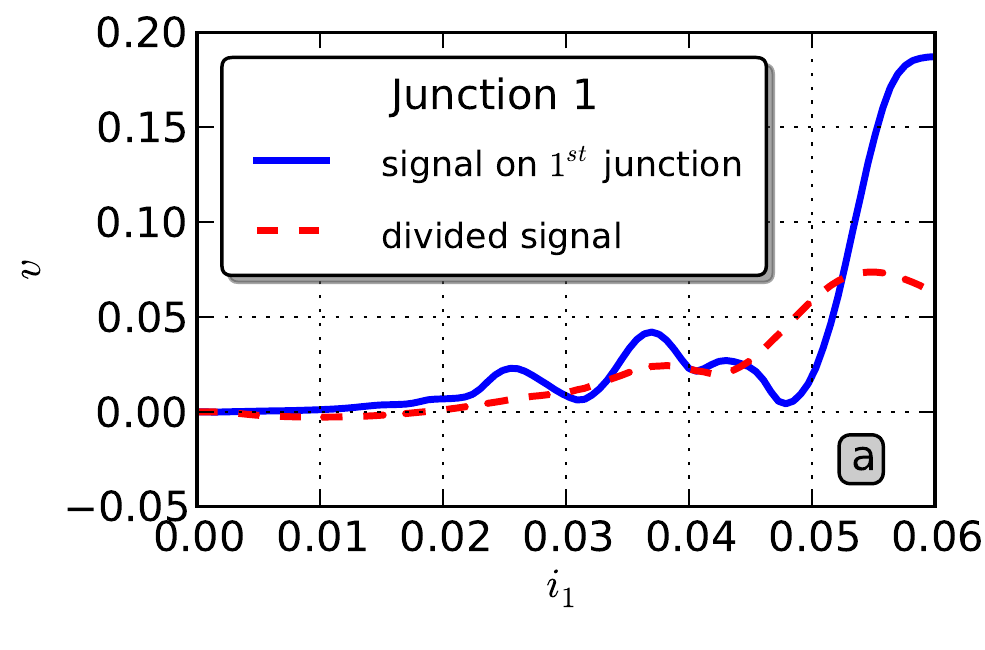}
	\includegraphics[width=0.49\textwidth]{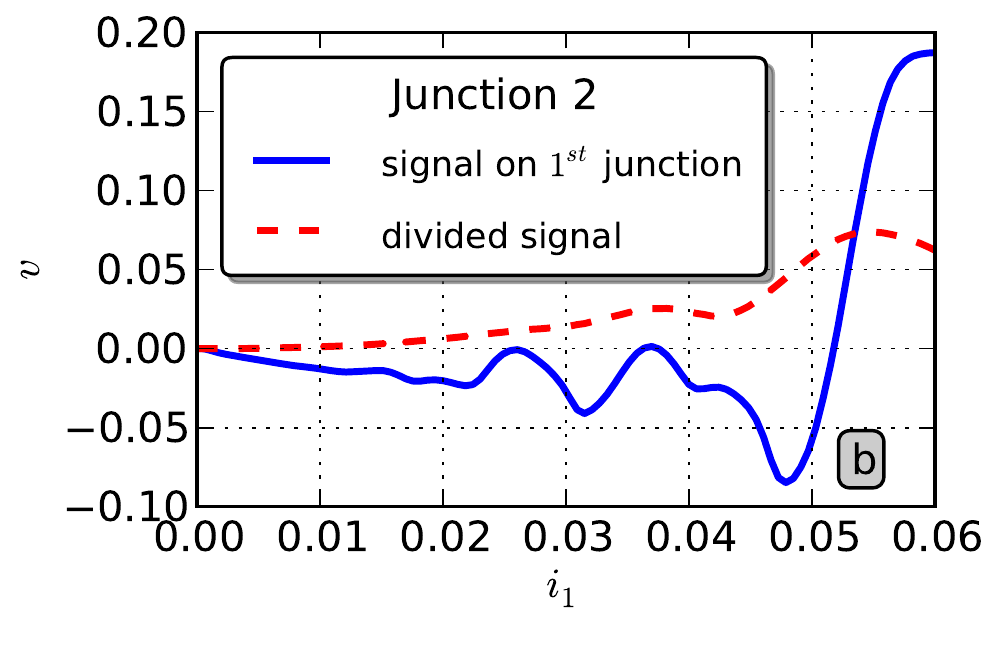}
	\caption{(color online). The second scenario: the long-time averaged voltage $v$ across the first (panel a) and the second 
	(panel b) junction. The panels show the dependence on the dc bias $I_{1}$ at fixed temperature 
	$D = 2 \cdot 10^{-5}$ for two different types of driving, ie. applied to only one junction 
	($a_{1} = 1.775, a_{2} = 0$, solid, blue line) and applied to both of them 
	($a_{1} = 0, a_{2} = 1.775$, dashed, red line). Other parameters are coupling strength $\alpha = 0.77$, 
	the frequency of the ac driving $\omega = 0.1875$.}
	\label{fig5}
\end{figure}
\begin{figure}[htbp]
	\centering
	\includegraphics[width=0.49\textwidth]{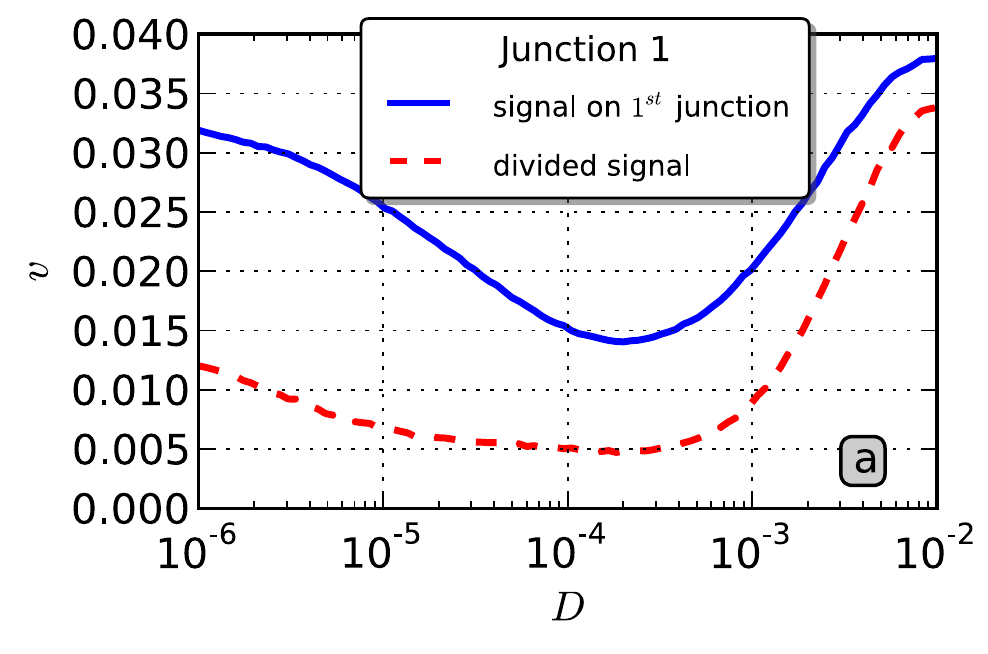}
	\includegraphics[width=0.49\textwidth]{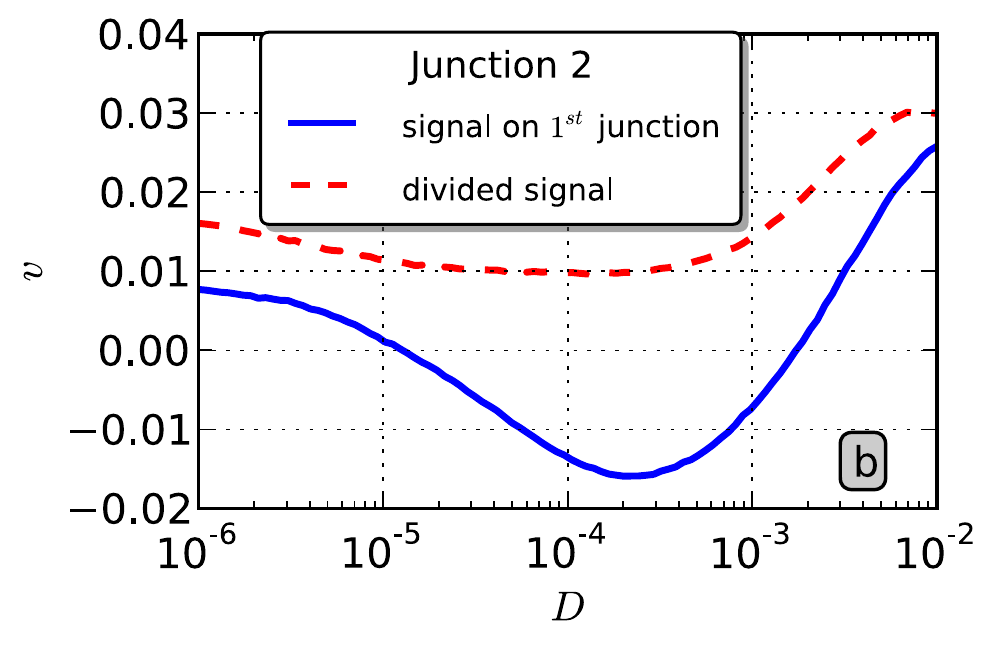}\\		
	\includegraphics[width=0.49\textwidth]{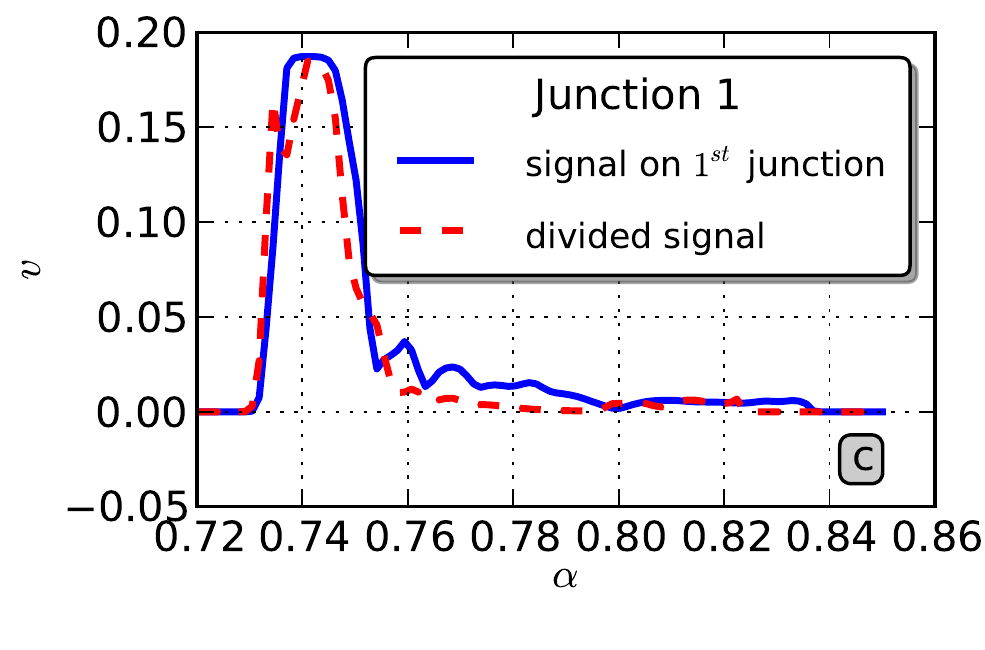}
	\includegraphics[width=0.49\textwidth]{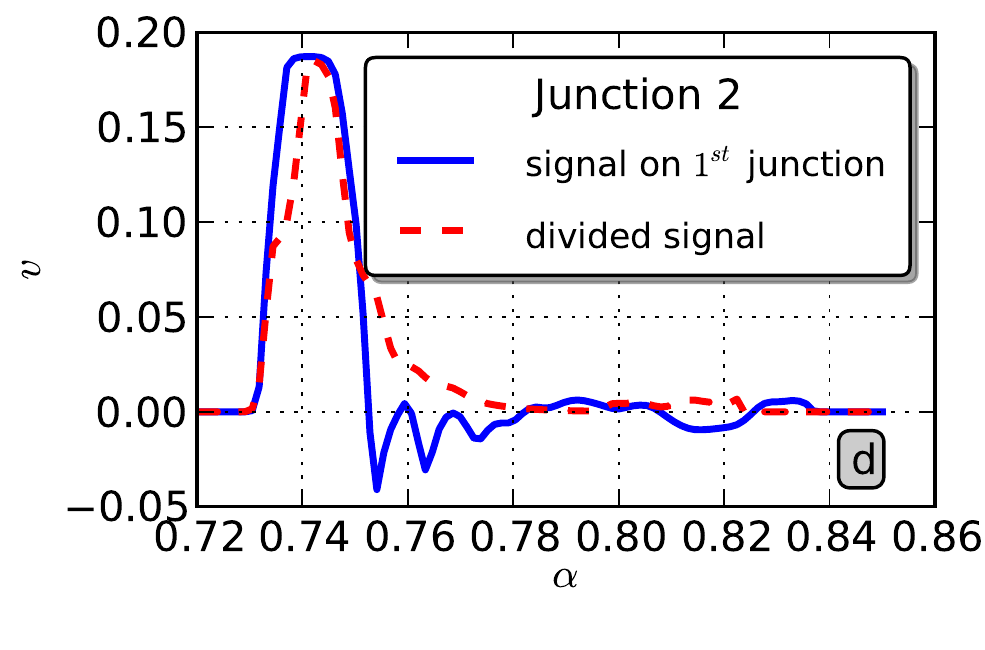}
	\caption{(color online). The long-time averaged voltage $v$ across the first (panels a and c) 
	and second (panels b and d) junction. 
	The upper panels show the dependence on temperature $D$ at fixed value of dc bias $I_{1} = 0.025$ 
	and the coupling strength $\alpha = 0.77$ for two different types of driving, i.e. applied to only 
	one junction ($a_{1} = 1.775, a_{2} = 0$, solid blue line) and applied to both of them 
	($a_{1} = 0, a_{2} = 1.775$, dashed red line). 
	The bottom panels show the role of the coupling $\alpha$ for   temperature $D = 2 \cdot 10^{-5}$.
	The frequency of the ac driving is $\omega = 0.1875$.}
	\label{fig6}
\end{figure}

 In the previous  two sections we studied properties of the DC voltage across the first and second 
 junctions driven by  two different  external currents. We presented the most interesting regimes 
 where anomalous transport (i.e. negative resistance) can occur. There are three  necessary 
 ingredients for the anomalous transport to observe: DC bias, AC current (the nonequilibrium driving) 
 and coupling.   In this section, we compare transport properties  in the same parameter domain but 
 for two scenarios.  In Fig. \ref{fig5}, the current-voltage curves are compared  in the regime where  
 the ANR and ANR is induced for the second junction in the first scenario and the DC voltage across  
 the first junction is always positive, cf. Fig. \ref{fig2}(a). On the other hand, in the second 
 scenario, the first junction exhibits very small  absolute negative resistance while the DC voltage 
 across  the second junction is always positive. 
It means that the unbiased AC current can change the direction of transport (of course, the DC 
current can change it but it is rather trivial because the DC current is biased).  In Fig. \ref{fig6}, 
we compare the above characteristics in dependence on temperature (the upper panels) and the coupling 
strength (the bottom panels) in the regime where anomalous transport is induced by thermal fluctuations. 
For  positive values of the DC current, one could expect four transport  regimes:
\begin{itemize}
\item[(I)] $v_1 > 0$ and $v_2 >0$,
\item[(II)] $v_1 > 0$ and $ v_2 <0$,
\item[(III)] $v_1 < 0$ and $ v_2 < 0$,
\item[(IV)] $v_1 < 0$ and $ v_2 > 0$.
\end{itemize}
In the first scenario, the regimes (I)-(III) can occur.  In the second scenario, 
the regimes (I) and (IV) can occur. From this point of view, the first scenario seems to be more optimal: 
there are three regimes. From the symmetry of the system it follows that the regime (IV) could  be obtained 
by applying the same driving to the second junction only. 
 
In summary, we studied  transport properties of two coupled  Josephson junctions and 
compared two scenarios for controlling the current-voltage characteristics when the system is driven 
by an external biased DC current and unbiased AC current consisting of one harmonic.  
We uncovered a reach diversity of anomalous  transport regimes for the first and second driving scenarios.

\section*{Acknowledgment} 
The work supported in part by the grant N202 052940 and  the  ESF Program  "Exploring the Physics of Small Devices".


\end{document}